%% file: apjl.tex
\documentclass[12pt,preprint]{aastex}

\newcommand{\kms}{~km\,s$^{-1}$}
\newcommand{\sna}{SN\,1987A}

\shorttitle{HETG observation of \sna}
\shortauthors{Dewey et al.}

\begin{document}

\title{Chandra HETG Spectra of \sna\ at 20 years}

\author{D. Dewey\altaffilmark{1}, S.A. Zhekov\altaffilmark{2,3},
R. McCray\altaffilmark{2}, C.R. Canizares\altaffilmark{1} }

\altaffiltext{1}{MIT Kavli Institute, Cambridge, MA 02139;
dd@space.mit.edu, crc@mit.edu}
\altaffiltext{2}{JILA, University of Colorado, Boulder, CO 80309-0440;
zhekovs@colorado.edu, dick@jila.colorado.edu}
\altaffiltext{3}{On leave from Space Research Institute, Sofia, Bulgaria}

\begin{abstract}

We have undertaken deep, high-resolution observations of
\sna\ at $\approx$\,20 years after its explosion with the Chandra HETG
and LETG spectrometers.  Here we present the HETG X-ray
spectra of \sna\ having unprecedented spectral resolution and
signal-to-noise in the 6\,\AA\ to 20\,\AA\ bandpass, which includes the
H-like and He-like lines of Si, Mg, Ne, as well as O\,VIII lines and
bright Fe\,XVII lines.
In joint analysis with LETG data,
we find that there has been a significant decrease from 2004 to 2007 in the average temperature
of the highest temperature component of the shocked-plasma emission.
Model fitting of the profiles of individual HETG lines
yields bulk kinematic velocities of the higher-Z ions, Mg and Si,
that are significantly lower than those inferred from the
LETG 2004 observations.

\end{abstract}





\section{Introduction}

Twenty years after outburst, \objectname[SN 1987A]{\sna}
is now well into its supernova remnant
phase, such that its observed luminosity is dominated by the impact of the
supernova debris with its circumstellar medium \citep{McCray2007}\footnote{See
also the ``Program'' link at: {\tt
http://astrophysics.gsfc.nasa.gov/conferences/supernova1987a/}}, the X-ray
luminosity has brightened by a factor $\sim 25$ since first
observed by Chandra in 1999 \citep{Aschenbach2007},
and is currently brightening at a rate $\sim 40
\%$ per year \citep{Park2007}.  The X-ray image seen by
Chandra is an expanding elliptical ring; its brightness distribution is
correlated with the rapidly brightening optical hotspots on the inner
circumstellar ring seen with the Hubble Space Telescope.  Assuming that the
X-ray emission is a circle inclined at $45^\circ$ (north side toward Earth),
its inferred radial velocity is
currently $1412 \pm 354 $\kms \citep{Park2007}.

\citet{Michael02} reported the first observation of the X-ray spectrum
taken with the High Energy Transmission Grating (HETG) on Chandra in October
1999.  That observation, of 116 ks duration, showed that the X-rays had an
emission line spectrum characteristic of shocked gas.  But that spectrum had
insufficient counting statistics (e.g., only 19 counts in O\,VIII L$\alpha$)
to provide reliable line profiles or accurate line ratios.  As the X-ray
luminosity of \sna\ has continued to increase, it has become possible to
measure emission line ratios and profiles with sufficient accuracy to permit
quantitative interpretation of the shock interaction responsible for the
X-ray emission.  Dispersed X-ray spectra of \sna\ have been
obtained with the Reflection Grating Spectrometer on XMM-Newton in May 2003
\citep{Haberl2006} and again in January 2007 \citep{Heng2008}, and with
the Low Energy Transmission Grating (LETG) on Chandra (289 ks duration) in
August-September 2004 (\citet{Zhekov05,Zhekov06} -- hereafter Z05, Z06).  

With angular resolution FWHM $\sim 0.7\arcsec$, Chandra can resolve the
circumstellar ring ($1.2\arcsec \times 1.6\arcsec$) of \sna.  This angular resolution
is vital for interpreting the linewidths seen in the dispersed X-ray
spectra, which are a convolution of the spatial structure of the X-ray image
and the kinematics of the X-ray emitting gas.  As Z05 have described, one
can separate these contributions by comparing the $m = +1$ and $m = -1$
orders of the dispersed spectrum: with the dispersion axis along the N-S
direction, the images of the ring in spectral orders dispersed to the
north will be ``compressed'', and those to the south will be ``stretched'' (Z05).  

\begin{figure*}[t]
\includegraphics[angle=270,scale=0.66]{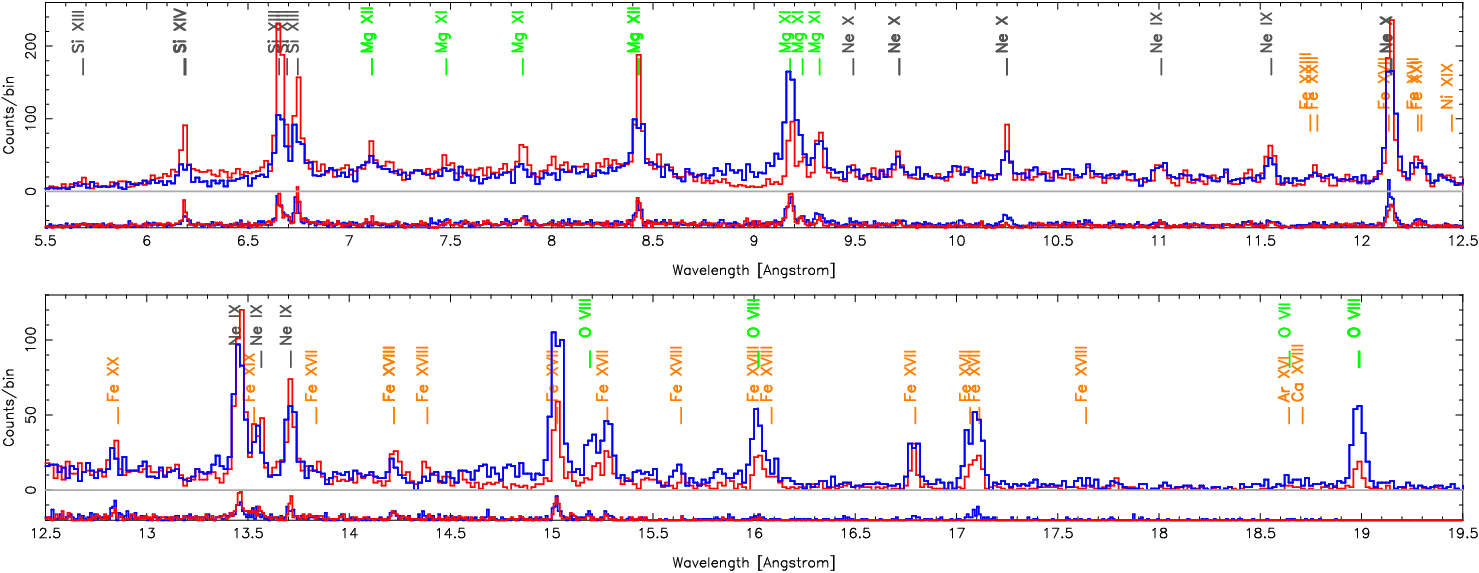}  \\
\caption{HETG \sna\ Spectra. Observed counts
spectra from all of the observations, totaling 360~ks, are shown; the
MEG minus (blue) and plus (red) order spectra are plotted along with offset
HEG spectra (same color coding.) The MEG[HEG] bin size is 0.020[0.010]\,\AA,
equivalent to $\approx$~1\arcsec\ in spatial extent.
The wavelengths of expected bright
lines are indicated as well. \label{fig:hiresspect}}
\end{figure*}

We have obtained new Chandra spectra of
\dataset[ADS/Sa.CXO#Obs/8523]{\sna, in March 2007 with the HETG}
(HETG'07) and in September 2007 with the LETG (LETG'07).  In this {\it
Letter} we focus on the HETG results, which have substantially greater
counting statistics and better spectral resolution than the LETG'04 spectra
(Z05), and which lead to a significant clarification of Z05's inferred
relationship between bulk motion and ion species.
Moreover, we see significant evolution of the
thermal structure of the interaction since the 2004 observations.
Preliminary analysis of the LETG'07 data
confirms that this evolution is not an effect of the spectrometer being
used.  In a subsequent paper, we will provide a more detailed analysis and
interpretation of the combined HETG and LETG observations.

\section{Observations and Data Reduction}

\begin{figure}[t]
\begin{center}
\includegraphics[angle=270,scale=0.60]{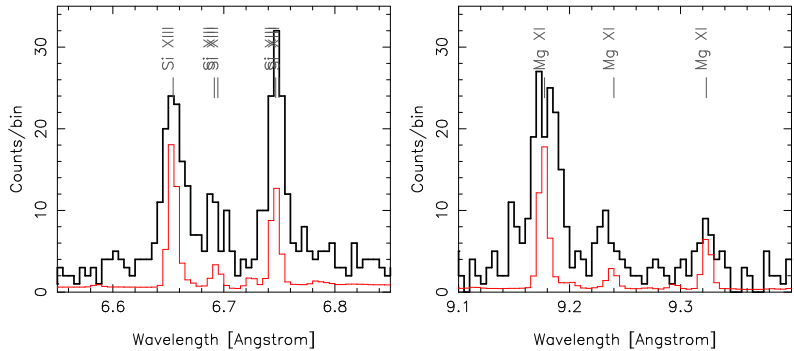}
\end{center}
\caption{Si and Mg triplets resolved.  The Si (left)
and Mg (right) triplets are plotted as observed in the plus (``compressed'') order
of the HEG.  The data are shown by the solid black histogram
and an arbitrarily-scaled point-source version of the 2-shock model
(no spatial-velocity effects)
is shown in red to suggest the relative similarity of the data and
model G-ratios ( (f+i)/r ). 
\label{fig:triplets}}
\end{figure}

The HETG'07 observations were carried out as part of the GTO program in spring
2007 near a roll angle of 270 degrees,
imaging the ``stretched'' orders on the ACIS-S1 BI chip
(which has the best low-energy sensitivity.)
The data consist of 14
observations with roll angles from $270^{\circ}$ to
$238^{\circ}$ during SN day 7321 to 7358, for a
total live time of 354.9~ks.
We used the standard CIAO tools and HETG extraction procedures to process
the data sets.  The {\tt tgdetect} routine located the zeroth order near the
``dark center'' of each event image to better than 0.1~arc second.  
The nominally extracted
first-order spectra
contain the following numbers of counts: 15.2\,k~(MEG-1), 14.0\,k~(MEG+1),
5.8\,k~(HEG-1) and 4.7\,k~(HEG+1).
In Figure~\ref{fig:hiresspect} we display all four first-order counts
spectra in the high S/N range, 5.5 to 19.5~\AA.  The higher resolution and lower
effective area of HEG compared to MEG spectra are evident,
e.g., at the Ne triplet $\sim 13.5$\AA;
we also display close-ups of the beautifully resolved He-like triplets
for Si and Mg in Fig.~\ref{fig:triplets}.

\clearpage

\section{Shock Emission Measure Distribution}
\label{sec:dem}

Following the analysis procedures of Z06,
we performed global fits to the data with two models: (i) a
simple 2-shock model and (ii) a distribution of shock emission
measures model, ``$EM(T)$''.  In the latter model, the
basic vectors characterizing the spectral distribution are plane-parallel
shocks characterized by their mean post-shock temperature
and a related ionization timescale.

We simultaneously fit the six background-subtracted $\pm1^{\rm st}$-order
spectra from the LETG'04, HETG'07 (MEG) and LETG'07 data sets,
rebinning all spectra to have a minimum of 30 counts per bin.  We assumed
common values among the data sets for $N_{\rm H}$, the interstellar absorption column density,
as well as the chemical composition of the hot, shocked plasma.
Specifically, we fixed the
abundances\,\footnote{We specify abundances relative to the AG89 values.
It is possible to use another reference set, however in that case
the ``fixed'' values must be converted to the new system to maintain
the same elemental number ratios.}
of H, He,
C, Ar, Ca and Ni to the values given in Z06, and we allowed the abundances
of N, O, Ne, Mg, Si, S and Fe to vary in our fits.

\begin{figure}[t]
\epsscale{0.66}
\plotone{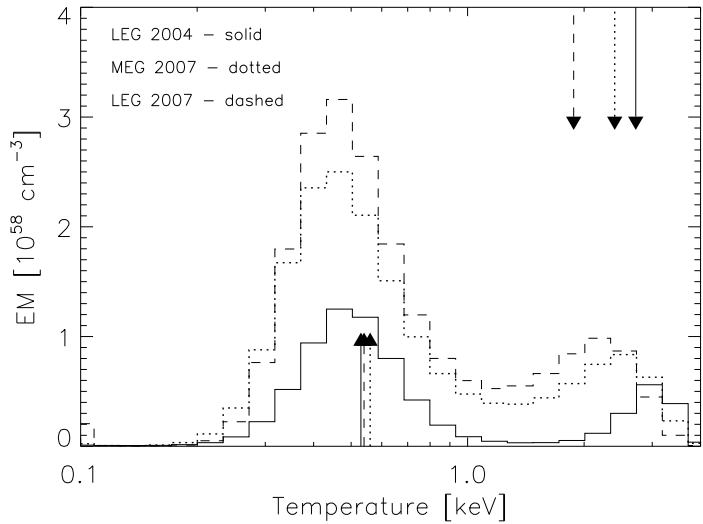}
\caption{Emission measure distribution versus shock temperature. $EM(T)$ as
derived with 25 points logarithmically spaced in the ($0.1-4$~keV)
post-shock temperature range.  A disproportionate growth of $EM(T)$ is seen in the
mid-T range, $\sim 1.4$~keV.  For reference, the arrows indicate the post-shock
temperatures of the simple 2-shock model fits given in the text;
low (high) temperature components are
plotted from the lower (upper) x-axis.
\label{fig:dem}}
\end{figure}

In principle some evolution of the chemical composition might take place as
a result of evaporation of dust grains in the shocked gas.  However, as
\citet{Dwek2007} have shown, the dust grains in \sna\ appear to have such
low abundance of that they cannot contribute substantially to the gas phase
abundance of heavy elements.  Likewise we are assuming uniformity of
composition throughout the ring.
These assumptions of stationary and uniform abundances in time and space
are taken as a starting point for these analyses.

For the 2-shock model joint fits,
the derived values and 90\,\% confidence limits (in brackets)
for the absorption column density
($N_{\rm H} = 1.30$ [1.18--1.46] $\times 10^{21}$)
and elemental abundances (N=0.56 [0.50--0.65],
O=0.081 [0.074--0.092], Ne=0.29 [0.27--0.31],
Mg=0.28 [0.26--0.29],
Si=0.33 [0.32--0.35], S=0.30 [0.24--0.36] and Fe=0.195 [0.189--0.206])
are all within or very near the 90\,\% limits presented in Z06.  
The values of $kT_{\rm low}$ (keV) are very similar for all three data sets:
0.53 [0.50--0.55], 0.56 [0.53--0.59], and 0.54 [0.53--0.56] for the
LETG'04, HETG'07, and LETG'07 data, respectively.
In contrast, the $kT_{\rm high}$ (keV) values
indicate a general evolution toward lower values across these same data sets:
2.7 [2.5--3.0], 2.4 [1.9--2.7], and 1.9 [1.8--2.0]\,\footnote{
For completeness the joint fit has $\chi^2=1819$ for 2248 degrees of freedom, and
the $\tau=n_e t$ values (in $10^{11}$~s\,cm$^{-3}$)
show an increasing (older, denser) trend ``as expected'': $\tau_{ T_{\rm low}}=$ 3.2, 3.6, and 4.8;
$\tau_{ T_{\rm high}}=$ 1.6, 2.2, and 2.7.}.
This result confirms a similar trend seen in the ``$kT({\rm hard)}$''
values derived from ACIS monitoring of \sna\ \citep{Park2006,Park2007}.

The decrease in $kT_{\rm high}$ is echoed by an evolution
of the shock emission measure distributions, $EM(T)$,
that we derive from global
fits to the three data sets, Figure~\ref{fig:dem}.
First, note that the fits to the
2007 LETG and MEG data are very similar. The value of $EM(T)$ derived from
LETG'07 data is slightly greater than that derived from the 2007 MEG
because the source brightened by a factor $\approx 1.17$ between Mar-Apr
2007 and Sep 2007.  We see more substantial evolution in $EM(T)$ from 2004
to 2007.  Like the 2004 $EM(T)$, the 2007 $EM(T)$ is bimodal, having a low
temperature peak at 0.55 keV and a high temperature peak at $\sim 2$ keV.
During the interval Sep 2004 to Mar-Apr 2007, the shape of the low
temperature peak did not change, but the total emission brightened by a
factor $\sim 2.47$,
corresponding to a doubling timescale of approximately two years.
At higher temperatures the shape of the $EM(T)$ appears to have evolved
substantially in addition to the increased amplitude.  We see substantial
``filling in'' of emission measure in the intermediate $T$ range,
1.0--2.5~keV,
and a decrease in the highest temperature bin ($>3$~keV).
Further analysis will be needed to quantify the significance
of these apparent changes.

\section{Line Profiles: Expanding Ring Model}
\label{sec:lines}

We fit the observed line profiles 
with a simple model incorporating the basic spatial-kinematic properties of
\sna's geometry, enabling us to infer an average expansion
velocity, $v_{\rm ring}$, for each emission line.
The geometric model is a torus
with specified inner (1.55\arcsec) and outer (1.7\arcsec) diameters with
its axis oriented in space ($42.8^\circ$~S and $10^\circ$~E from the
line-of-sight.)  The torus has a non-uniform
azimuthal intensity distribution
determined from the observed zeroth-order imaging data.
Emission at each location on the ring is Doppler shifted by the
combination of the component of radial expansion at that point along the line of sight,
the overall systemic velocity, and 
an additional Gaussian Doppler broadening term, $\Delta v_{\rm los}$, along the
line-of-sight.

We implemented the above as a custom model\,\footnote{The convenient
creation and usage of 3D geometric models in astrophysical data analysis is
one component of the Hydra project at MIT, see, e.g., {\tt
http://space.mit.edu/hydra/v3d.html}}
in ISIS~\citep{HouckISIS}.
We use a Monte Carlo scheme to evaluate the line profile
for the particular model parameters, also taking into account the
properties of the specific data set (the grating,
order, and observation roll angle.)
We sum one (or two in the case of blended lines)
of these ``ring model'' line shapes with the nominal 2-shock model in which the
abundance of the element being fit is set to zero.  We tie the 2-shock norms
together in the appropriate ratio and allow them to vary in
the line fitting.  In this way, we have an accurate continuum shape and also make some
allowance for contaminating lines, e.g., Fe lines in the Ne triplet region.

\begin{figure}[t]
\begin{center}
\includegraphics[angle=270,scale=0.60]{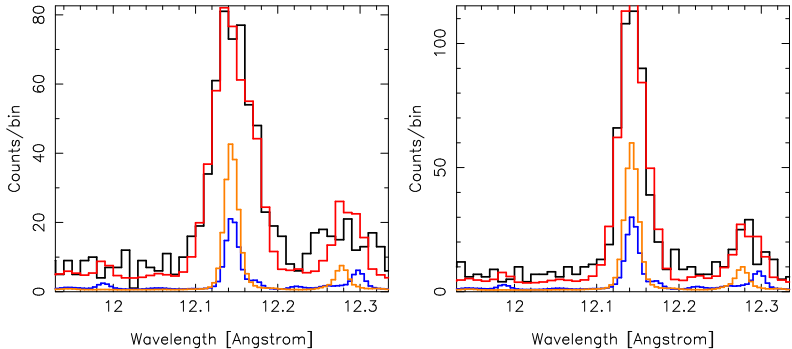}
\end{center}
\caption{Example of the spatial-spectral model fit.
The MEG minus (left) and plus (right) order data (black)
for the Ne\,X line are show with the
best-fit model overplotted (red.)
For reference, the low and high $T$ 
components of the  2-shock model are shown at point-source resolution (orange and
blue, respectively.)
\label{fig:ssfit}}
\end{figure}

\begin{figure}[h]
\begin{center}
\includegraphics[angle=0,scale=0.66]{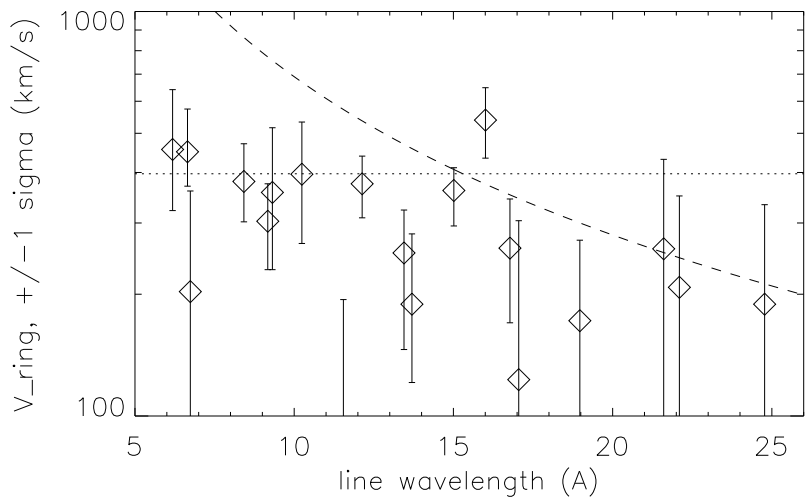}
\end{center}
\caption{``Ring'' velocities versus wavelength.  The HETG-measured 
radial velocities and their one sigma ranges are plotted against the line wavelength.
For comparison the ``stratified'' (dashed) and ``constant'' (dotted)
radial velocities from Z05 are overplotted.  The stratified model is clearly
excluded by the new HETG data.
\label{fig:linevels}}
\end{figure}

We fit the data from the first 319~ks (excluding the last observation
taken at a roll angle $25^\circ$ from the average) with this
model in a limited wavelength region around each line.
We used all four HEG/MEG $\pm1$-order spectra except in
the long wavelength range ($> 16$\AA) where we excluded one or both
of the HEG spectra from the fits because of a lack of counts.
Figure~\ref{fig:ssfit} illustrates
the fit to the Ne~X line in the MEG $\pm1$-order spectra. 
We tabulate the results for the bright lines in the spectrum in Table~\ref{tbl-lines}
and display their ring expansion velocities in Fig.~\ref{fig:linevels}.
The best-fit values for the additional Doppler broadening parameter,
$\Delta v_{\rm los}$, (not tabulated)
are in the 300--700\kms\ FWHM range.

\clearpage

\input{tab1.tex}

\clearpage

The bulk velocities measured with the HETG'07 data
(the diamond symbols in Figure~\ref{fig:linevels}) are
significantly lower, by a factor of 2 at Si and Mg, than those inferred from the
``stratified'' model that was employed to fit the LETG'04 FWHM data
(dashed line).  The new HETG data 
are, however, in good agreement with the constant velocity model of Z05
(dotted line).
Note that there is the hint of a much reduced ``stratified'' trend:
expansion velocities tending to increase toward shorter wavelengths.

\vfill

\vfill

\clearpage

\section{Discussion}
\label{sec:discuss}

The most striking result that we find from the HETG data is the relatively
low bulk radial velocity of the shocked gas in the ring.
In a simple model consisting of a plane-parallel strong shock entering a 
stationary gas, the bulk velocity of the shocked thermal plasma should be 
given by $v_{\rm psh} = 3/4 \, v_{\rm sh}$, and
the post-shock mean plasma temperature should be given by $kT =
1.4(v_{\rm sh}/1000~\mathrm{km\,s^{-1}})^2$~keV (for mean
molecular weight $\mu=0.72$ based on our abundances, see also Z06).
Then, setting $v_{\rm psh} = v_{\rm ring}$, we
would expect post-shock temperatures in the range $kT \sim 0.1-0.6$~keV
for the observed range $v_{\rm ring} \sim 200-500$\kms\ 
(Fig.~\ref{fig:linevels}).  This temperature range is much lower than the
range $0.3-3$~keV inferred from the spectral modeling of the emitting gas.
Clearly, such a simple shock model is inadequate to describe the actual
system.  

The range of broadening, $\Delta v_{\rm los}$, used in the fits to the line
profile, 300-700\kms\ FWHM, generally exceeds the expected thermal
line-of-sight contribution,
given by $\Delta v_{\rm therm} = 163\,\sqrt{E_{\rm keV}/(A/20)}$\kms~FWHM
where $E_{\rm keV} = kT_{\rm i}$ is the ion temperature in keV and $A$ is the
atomic weight of the ion.  For $E_{\rm keV}=0.4-3$ the range of thermal Doppler
widths (FWHM) is: 115--315\,(O); 87--237\,(Si); and 61--170\,(Fe).

The fact that the X-ray image is correlated with the optical hotspots leads
us to a picture in which the blast wave ahead of the supernova debris is
overtaking dense clumps of circumstellar gas associated with the hotspots.
Since these hotspots are unresolved, we can only guess at their geometry and
density distribution.  But it would be reasonable to expect that the X-ray
emitting gas spans a range of densities and has complex morphology.

If a blast wave runs into a clump of high density at {\it normal incidence}, the
transmitted shock might be too slow to emit X-rays but the shock
encountering the clump would be reflected.  The reflected shock would leave
behind twice-shocked gas having nearly stationary bulk velocity but further
elevated temperature.  As more and more of the X-ray emission comes from gas
behind such reflected shocks, we would expect that the fraction of the X-ray
emission measure at higher temperatures would increase, while the average bulk
velocity would decrease.  And that is what we see.

In addition, much of the X-ray emission probably comes from shocks resulting from the
blast wave encountering dense clumps at {\it oblique incidence}, in which case the
shocked gas would have significant velocity components parallel to the shock
surface.  We suspect that the complex hydrodynamics resulting from both
transmitted and reflected shocks encountering the circumstellar ring at
normal and oblique incidence is responsible for the Doppler broadening seen
in the line profiles.

As Table 1 shows, the radial expansion velocity of the ring inferred from
the X-ray line profiles ranges from $\sim200-450$\kms, much less than
the value $1412\,\pm354$\kms\ inferred from the expansion of the
X-ray image \citep{Park2007}.  The expansion of the X-ray image tells us
the location of the centroid of the X-ray emitting ring, which is determined
by the average radius of the relatively dense ($n > 10^4$ cm$^{-3}$) shocked
gas.  We believe that these dense clumps or fingers are being overtaken by a
blast wave that is propagating at radial velocities $v > 2000$\kms\ through
gas of relatively low density ($n < 10^2$ cm$^{-3}$), which does not
contribute substantially to the observed X-ray emission.  On the other hand,
the velocities seen in the X-ray emission line profiles represent the actual
kinematic velocities of the shocked gas surrounding the dense clumps, which are much
less than the velocity of the blast wave. 

There is much more that can be done with this data set -- in particular
looking at the full 2D distribution of the dispersed events instead of just
their 1D projection.  Especially in the ``stretched'' MEG-minus order, one
may be able to measure spatially-resolved line ratios.  Given the complexity
of the \sna\ system, such spatial analysis may identify emission from
other geometric components of the system.

\acknowledgments
Support for this work was provided by the National Aeronautics and
Space Administration through the Smithsonian Astrophysical Observatory
contract SV3-73016 to MIT for Support of the Chandra X-Ray Center,
which is operated by the Smithsonian Astrophysical Observatory for and
on behalf of the National Aeronautics Space Administration under contract
NAS8-03060, and by Chandra grant GO7-8062X to the University of Colorado.

{\it Facilities:} \facility{CXO (HETG)}.

\end{document}

%% file: tab1.tex
%
%

\begin{deluxetable}{lrrll}
\tablewidth{3.5in}
\tabletypesize{\scriptsize}
\tablecaption{Fit Parameters for the Non-uniform Ring Model. \label{tbl-lines}}
\tablehead{
\colhead{Ion} & \colhead{$\lambda_{\rm t}$\tablenotemark{\,a}} &
\colhead{$\lambda_{\rm fit}$\tablenotemark{\,b}$\,(1\sigma)$} &
\colhead{$v_{\rm ring}\tablenotemark{\,c}\,\pm1\,\sigma$} & 
\colhead{Flux\tablenotemark{\,d}}  
}
\startdata
Si\,XIV L$\alpha$  &  6.183  &  6.181(1)  & 456.,\ 322--641 &
\ \,4.6~${{\pm10\%}}$ \\
Si\,XIII r~(\tablenotemark{\,e})         &  6.648  &  6.650(1)  & 450.,\ 370--574 &
14.5~${{\pm\ 5\%}}$ \\
\ \&\ Si XIII i   &  6.687  &  6.691(5)  & \ '' &
\ \,3.4~${{\pm16\%}}$ \\
Si\,XIII f         &  6.740  &  6.738(1)  & 203.,\ \ \ \,0--360 &
\ \,6.3~${{\pm\ 9\%}}$ \\
Mg\,XII L$\alpha$  &  8.422  &  8.421(1)  & 380.,\ 302--471 &
11.0~${{\pm\ 6\%}}$ \\
Mg\,XI r~(\tablenotemark{\,e})           &  9.169  &  9.170(1)  & 303.,\ 230--375 &
22.3~${{\pm\ 5\%}}$ \\
\ \&\ Mg\,XI i     &  9.230  &  9.228(3)  & \ '' &
\ \,5.3~${{\pm15\%}}$ \\
Mg\,XI f           &  9.314  &  9.315(2)  & 357.,\ 230--516 &
\ \,7.0~${{\pm12\%}}$ \\
Ne\,X L$\beta$     & 10.239  & 10.241(2)  & 396.,\ 267--533 &
10.8~${{\pm10\%}}$ \\
Ne\,IX  3-1        & 11.544  & 11.545(2)  &\ 32.,\ \ \ \ 0--194 &
11.0~${{\pm12\%}}$ \\
Ne\,X L$\alpha$    & 12.135  & 12.134(1)  & 375.,\ 309--439 &
72.0~${{\pm\ 4\%}}$ \\
Ne\,IX r~(\tablenotemark{\,e})           & 13.447  & 13.448(1)  & 253.,\ 146--323 &
62.0~${{\pm\ 6\%}}$ \\
\ \&\ Ne\,IX i     & 13.552  & 13.552(3)  & \ '' &
16.0~${{\pm16\%}}$ \\
Ne\,IX f           & 13.699  & 13.698(1)  & 189.,\ 121--282 &
36.7~${{\pm\ 9\%}}$ \\
Fe\,XVII           & 15.014  & 15.014(1)  & 361.,\ 295--411 &
75.0~${{\pm\ 6\%}}$ \\
O\,VIII L$\beta$   & 16.006  & 16.006(3)  & 539.,\ 434--648 &
42.6~${{\pm11\%}}$ \\
Fe\,XVII           & 16.780  & 16.775(3)  & 260.,\ 170--344 &
33.5~${{\pm13\%}}$ \\
Fe\,XVII~(\tablenotemark{\,e})           & 17.051  & 17.051(4)  & 123 ,\ \ 60--304 &
27.4~${{\pm17\%}}$ \\
\ \&\ Fe\,XVII     & 17.096  & 17.094(3)  & \ '' &
41.7~${{\pm13\%}}$ \\
O\,VIII L$\alpha$  & 18.970  & 18.970(3)  & 172.,\ 70--272 &
103.~${{\pm10\%}}$ \\
O\,VII r           & 21.602  & 21.60(1)  & 259.,\ \ \ \,0--431 &
\ \,28.~${{\pm40\%}}$ \\
O\,VII f           & 22.098  & 22.01(1)  & 208.,\ \ \ \,0--350 &
\ \,33.~${{\pm40\%}}$ \\
N\,VII L$\alpha$   & 24.782  & 24.77(1)  & 189.,\ \ 45--333 &
\ \,75.~${{\pm22\%}}$ \\
\enddata
\tablenotetext{a}{Theoretical wavelengths in \AA, from \citet{Huenem06}.}
\tablenotetext{b}{Measured wavelengths in \AA\ with last-digit error in (\,)'s.
Doppler and systematic (286.5\kms) velocities are then included
in the model.}
\tablenotetext{c}{This is the model radial velocity, in \kms, in the equatorial plane
at the average ring radius and its $\pm 1\,\sigma$ range.}
\tablenotetext{d}{Observed flux in $10^{-6}$~photons~cm$^{-2}$\,s$^{-1}$.}
\tablenotetext{e}{Two lines were jointly fit for this closely-spaced pair.}
%
\end{deluxetable}